\documentclass[twocolumn,english,10pt,showkeys,aps]{revtex4-2}

\pdfoutput=1

\usepackage{babel}
\usepackage{graphicx}
\usepackage{subfigure}
\usepackage{booktabs}  

\begin{document}

\makeatletter
\makeatother

\title{Featuring ACE2 binding SARS-CoV and SARS-CoV-2 through a conserved 
   evolutionary pattern of amino acid residues}

\author{Patr\'icia P. D. Carvalho}
\email{patricia.duzi.2018@usp.br}

\author{Nelson A. Alves}
\email{alves@ffclrp.usp.br}
\affiliation{Departamento de F\'{\i}sica, FFCLRP,
Universidade de S\~ao Paulo, \\
Avenida Bandeirantes, 3900. Ribeir\~ao Preto 14040-901, SP, Brazil.}

\begin{abstract}
    Spike (S) glycoproteins mediate the coronavirus entry into the host cell. 
The S1 subunit of S-proteins contains the receptor-binding 
domain (RBD) that is able to recognize different host receptors, 
highlighting its remarkable capacity to adapt to their hosts along the viral 
evolution. 
  While RBD in spike proteins is determinant for the virus-receptor
interaction, the active residues lie at the receptor-binding motif (RBM),
a region located in RBD that plays a fundamental role binding the outer surface 
of their receptors.
  Here, we address the hypothesis that SARS-CoV and SARS-CoV-2 strains able to 
use angiotensin-converting enzyme 2 (ACE2) proteins have adapted their RBM 
along the viral evolution to explore specific conformational topology 
driven by the residues YGF to infect host cells.
   We also speculate that this YGF-based mechanism can act as a 
protein signature located at the RBM to distinguish coronaviruses
able to use ACE2 as a cell entry receptor.
\end{abstract}


\maketitle


\section{Introduction}

   Viruses are the most numerous type of biological entity on Earth
and the identification of novel viruses continues to enlarge the known 
viral biosphere \cite{pan-2017,shi-2018}.
   This collection of all viruses presents enormous morphological and 
genomic diversity as a result of continuous exchange of genetic material 
with the host cells \cite{nasir-2014,koonin-2020}.
    Moreover, this well succeeded long-term virus-host interaction indicates
that viruses are more than simple genomic parasites in all cellular life forms
\cite{claverie-2006}.
    A number of evidences has led to the proposal that viruses play an
astonishing role as agents of evolution because of their capacity in propagating 
between biomes \cite{sano-2004} and in gene transfer between species
\cite{blerkom-2003,filee-2003,koonin-2013,enard-2016}.
    For this purpose, viruses have developed large number of genome replication 
and protein expression strategies to benefit from the host translational 
machinary over time \cite{barana-2003}.

    Despite all of such enormous diversity in gene sequence, 
it is not possible to achieve huge number of highly distinct
protein structures mainly because of stereochemical constraints 
on the possible protein folds \cite{weng-2020}.
    In fact, it has been observed common secondary structures throughout 
different virus families while the sequences are not fully conserved 
\cite{weng-2020,ahola-2015}.
    This may result in evolutionary efficiency once viruses can
exploit already well designed motifs from
similar cellular functions \cite{barana-2003}.
    
   Currently, the world population is confronting a new coronavirus disease 
(COVID-19), a highly infectious disease to humans.
   This disease is caused by severe acute respiratory syndrome coronavirus 2
(SARS-CoV-2) and is affecting human health worldwide.  
   Coronaviruses (CoVs) belong to the large and diverse family 
{\it Coronaviridae}, within the order 
{\it Nidovirales}
and suborder {\it Cornidovirineae} \cite{ICTV-2020}.
   Their subfamily {\it Orthocoronavirinae} contains four genera based
on phylogeny and termed as {$\alpha$},  {$\beta$}, {$\gamma$}, and
\mbox{$\delta$-{\it coronavirus}}.

   SARS-CoV-2 belongs to the $\beta$-{\it coronavirus} genus as well as
SARS-CoV, middle east respiratory syndrome coronavirus (MERS-CoV),
and hCoV-HKU1, to cite a few  \cite{jaimes-2020}.
   Other important representative human viruses as hCoV-NL63
and hCoV-229E belong to $\alpha$-{\it coronavirus}.
   Phylogenetic relationships among the known members of this subfamily
indicate that $\alpha$ and $\beta$-{\it coronavirus} infect mammals, while 
$\gamma$ and $\delta$-{\it coronavirus} infect both mammalians and avians.

   Members of {\it Coronaviridae} family are enveloped, positive 
single-stranded RNA (+ssRNA) viruses and render the largest genomes among 
all known RNA viruses \cite{masters-2006,cheng-2007,su-2016,li-2016-review}.
   The +ssRNA genomes undergo rapid mutational changes 
   \cite{sanjuan-2016},
leading to faster adaptation to new hosts,
though they also contain conserved sequence motifs as observed, for example,
in multiple alignments do CoV strains \cite{ahola-2015,lau-2007,woo-2012}.

   Coronaviruses attach to host cell surface receptors via their spike (S)
glycoproteins, located on the viral envelope, to mediate the entry 
into the host cell.
   Each monomer of trimeric S-protein comprises two subunits
S1 and S2, responsible for the viral attachment and for the membrane fusion,
respectively  \cite{beniac-2006,li-2012,song-2018}.
  The S1 coronavirus subunit contains the receptor-binding domain (RBD)
that is able to recognize different host receptors,
highlighting its remarkable capacity to adapt to their hosts along
the viral evolution.
   Thus, it is not unexpected to observe in this domain 
high sequence divergence even for the same coronavirus identified in 
different host species.
   In contrast, the S2 subunit presents the most conserved region 
in the S-protein.
   
    The binding of RBD spike proteins to the receptor on the host cell 
is the first step in virus infection. 
    This initial step is followed by an entry mechanism of enveloped viruses
into target cells.
     Usually, most viruses enter cells through endocytotic pathways with 
 the fusion occurring in the endosomes, although a direct entry into cells 
 can occur by fusion of their envelopes with the cell membrane  
 \cite{belouzard-2012}.

   A number of CoVs utilizes angiotensin-converting enzyme 2 (ACE2)
as the entry receptor into cells, exemplified by 
$\beta$-genus human respiratory SARS-CoV, SARS-CoV-2, and 
$\alpha$-genus hCoV-NL63
 \cite{jaimes-2020,hoffmann-2020,letko-2020,walls-2020}.
     In particular,  SARS-CoV, as well as SARS-CoV-2, enter the 
cell via endocytosis induced by RBD complexed with human ACE2 (hACE2) receptor
\cite{wang-2008-cell,wang-2008-virus,yuan-2017-nature,milewska-2018,ou-2020}.
     In contrast, the $\beta$-genus MERS-CoV and its genetically related
bat CoV-HKU4 utilize dipeptidyl peptidase 4 (DPP4) as the viral receptor 
\cite{yang-2014-PNAS}.
    Other viral receptor is aminopeptidase N (APN), recognized for example by 
the $\alpha$-genus hCoV-229E \cite{li-2015-review}.

    The human coronaviruses hCoV-HKU1, hCoV-229E, hCoV-NL63, 
and hCoV-OC43, 
cause mild to moderate upper respiratory tract infections 
\cite{weiss-2005}, while SARS-CoV and SARS-CoV-2 cause severe respiratory diseases, 
with SARS-CoV-2 being far more lethal than SARS-CoV. 
    SARS-CoV strains vary enormously in infectivity, which can be connected to
their binding affinities to hACE2 \cite{cui-2019-nature}.
    This binding affinity, in turn, can be correlated with disease severity 
in humans  \cite{li-2005-embo}. 

     While RBD in spike proteins is determinant for the virus-receptor
interaction, the active residues lie at the receptor-binding motif (RBM),
which is part of RBD and plays a fundamental role binding the outer surface 
of their receptors
\cite{hoffmann-2020,letko-2020,cui-2019-nature,wan-2020,li-2005-science309}.
   The importance of the RBM is further explored here in relation to
its structural topology.
    Thus,  instead of only analysing specific residues that make contacts with
ACE2 after binding, we go a step further and track the molecular origin that drives 
the viral attachment to this cell receptor.
    This investigation has revealed a highly conserved amino acid residue
sequence Tyr-Gly-Phe (YGF) in coronavirus variants that employ this receptor.
    Consequently, we hypothesize that the short sequence YGF is vital for RBD-ACE2
interaction because of the formation of a hydrophobic pocket proper to 
the receptor specificity \cite{wan-2020,wu-2012,lan-2020-nature,he-2020}.
    Moreover, we recognize that a similar binding mechanism is characteristic of the 
interaction between ubiquitin-associated (UBA) domain proteins and ubiquitin.
    In this vein, we conclude that is plausible that SARS-CoV and SARS-CoV-2 strains able 
to use ACE2 proteins have adapted their RBM along the viral evolution to explore
such a mechanism to infect host cells.

\begin{figure*}
\centering
\begin{minipage}{.47\textwidth}
  \centering
  \includegraphics[angle=0,width=9.3cm]{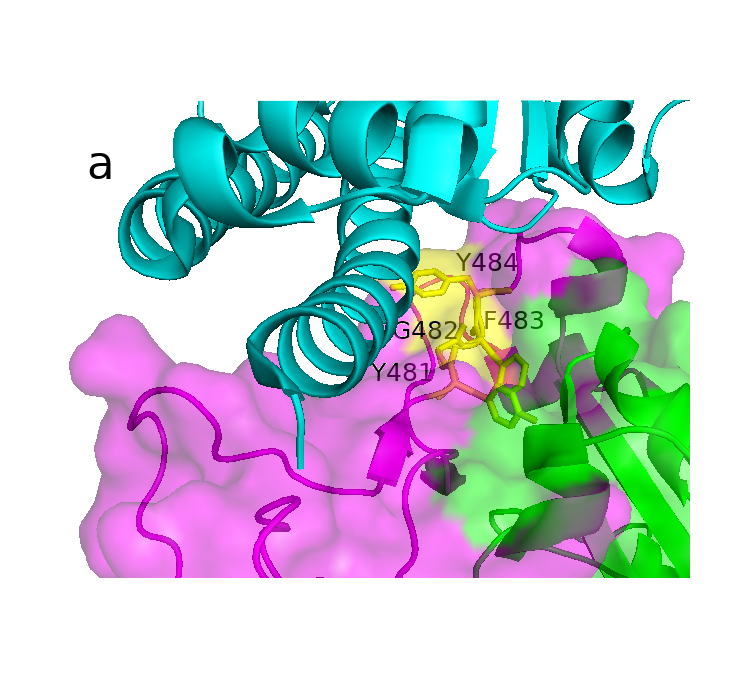}
\end{minipage}%
\hspace{0.1cm}
\begin{minipage}{.47\textwidth}
  \centering
  \includegraphics[angle=0,width=9.3cm]{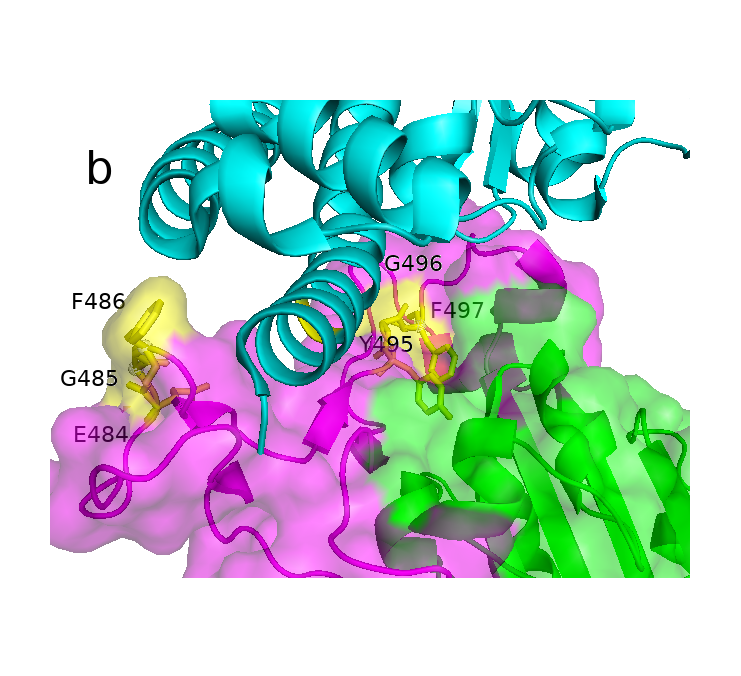}
\end{minipage}
\vspace{-0.3cm}
\caption{Detailed surface view of SARS-CoV and SARS-CoV-2 RBD.
(a) Residues YGFY at the interface of SARS-CoV complexed with  
hACE2 (PDB ID: 2AJF). 
    These residues are in yellow color and form a hydrophobic pocket 
located in the RBM (magenta color).
(b) Residues YGF and EGF (yellow color) at the interface of SARS-CoV-2 
 complexed with  hACE2 (PDB ID: 6LZG). 
  The first sequence is located in a hydrophobic pocket, while the
  second sequence EGF is on the RBM surface (magenta color).
   Ribbon representation of ACE2 is in blue color.}
   \label{fig:pocket_sars}
\end{figure*}

\subsection{The conserved XGF loop in UBA-ubiquitin interaction}

    Amino acid sequences of type XGF, where the residue  X is frequently the
residue Met, form a highly conserved loop characteristic of ubiquitin-associated 
(UBA) domain that occurs in a variety of proteins.
    The UBA domain is a conserved motif through eukaryotic evolution
and is found in many proteins related to the ubiquitin metabolism and in particular, 
associated with ubiquitin-mediated proteolysis
\cite{pickart-2004,finley-2009}.
    The MGF loop in the UBA domain is typical of a hydrophobic pocket
that is critical for recognition and binding affinity to ubiquitin through a 
hydrophobic surface patch located in the vicinity of this loop
\cite{dieckmann-1998,wilkinson-2001,bertolaet-2001,madura-2002,raasi-2005,long-2008,tse-2011,cabe-2018}.
    UBA domains are ubiquitin receptors whose binding is a fundamental step 
for diverse regulatory functions.
   
   NMR analyses of UBA-ubiquitin interactions identify hydrophobic
surface patches formed by the conserved MGF sequence as the main 
determinants for the protein-protein interaction.
   A number of alignments of UBA domains has revealed mutations
in the MGF sequence, mainly 
M$\rightarrow$L, M$\rightarrow$Q, and F$\rightarrow$Y, but these mutations still 
maintain the overall hydrophobic characteristic for the main set of residues located
at the interface on these UBA domains \cite{geetha-2002,mueller-2002}.

\begin{figure*}
\centering
\begin{minipage}{.47\textwidth}
  \centering
  \includegraphics[angle=0,width=8.8cm]{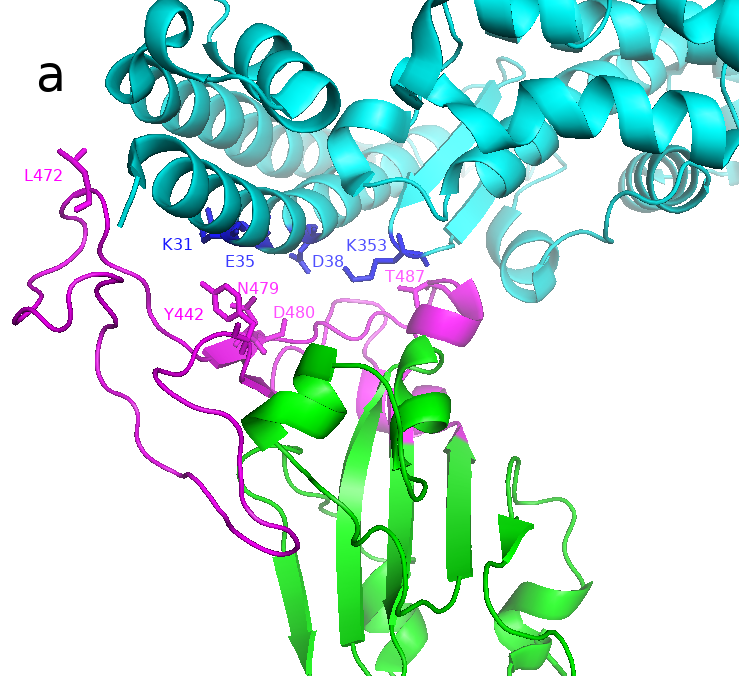}
\end{minipage}%
\hspace{0.2cm}
\begin{minipage}{.47\textwidth}
  \centering
  \includegraphics[angle=0,width=8.8cm]{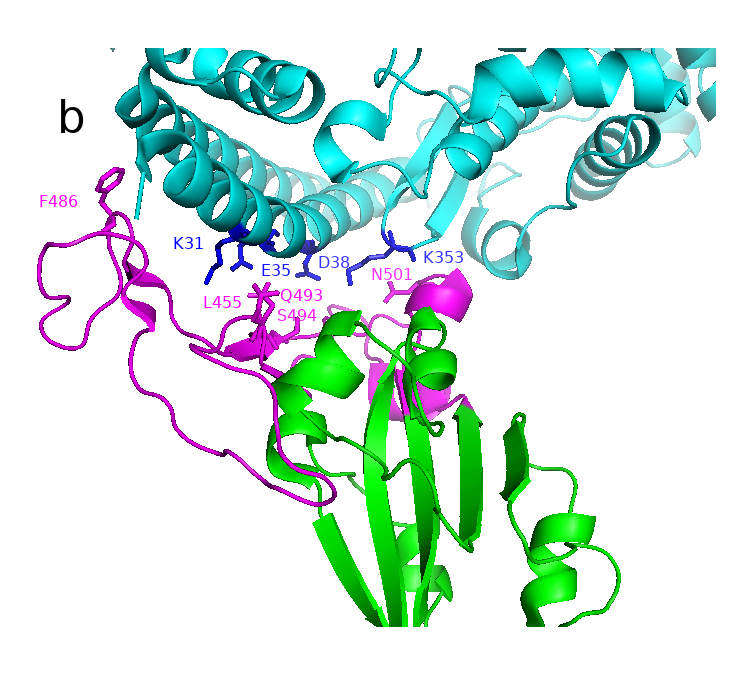}
\end{minipage}
\vspace{0.2cm}
\caption{SARS-CoV and SARS-CoV-2/hACE2 RBD interfaces.
   Ribbon diagrams of SARS-CoV RBD (a) and SARS-CoV-2 RBD (b) 
 complexed with hACE2 (blue color), where the 
 RBM is highlighted in magenta color.
  The main residues responsible for the structural binding are 
  displayed in the stick representation.}
 \label{fig:residues_sars}
\end{figure*}

\section{Results and Discussion}

\subsection{Spike receptor-binding motifs in human CoVs}

     Here, we investigate the occurrence and importance of the 
specific amino acid residue sequence YGF for SARS-CoV and SARS-CoV-2 
strains able to use ACE2 proteins as receptors.
     It is displayed in 
     Fig~\ref{fig:pocket_sars}a
the interface of SARS-CoV RBD spike-protein 
(magenta and green color) complexed with hACE2 (blue color) to gain insight 
about the importance of this type of conformational mechanism in creating a 
shape complementarity between receptor and ligand. 
    The RBM is in magenta color, with the yellow color displaying the 
YGFY sequence in that pocket, which establishes the proper
relative position for favorable binding to surface-exposed hACE2 residues.
    The YGFY sequence seems strongly conserved in SARS-CoV spike RBD,
more precisely located at residues 481-484 in the receptor binding motif.
    Noteworthy, this sequence seems to be unique because even the shorter  
YGF sequence does not occur in this region, neither in  the RBD.
    As a consequence of this hydrophobic pocket, amino acid residues responsible 
for binding interaction are located close to this conformational structure
as, for example, the residues N479 and T487 (Fig~\ref{fig:residues_sars}a).
  These residues have been identified to be essential for 
  SARS-CoV spike RBD/ACE2 binding 
\cite{li-2005-embo,li-2005-science309,qu-2005}.
    The residue N479 in SARS-CoV is located near K31 of hACE2 which in turn makes a 
salt bridge with E35, a residue buried in that hydrophobic environment.
   The residue  T487 is located close to K353 on hACE2, and in turn makes a
salt bridge with D38, also buried in that pocket.
    Other important residues for this attachment are Y442, L472, and D480
\cite{wan-2020}.

\begin{figure*}
\centering
\begin{minipage}{.48\textwidth}
   \centering
   \includegraphics[angle=0,width=9cm]{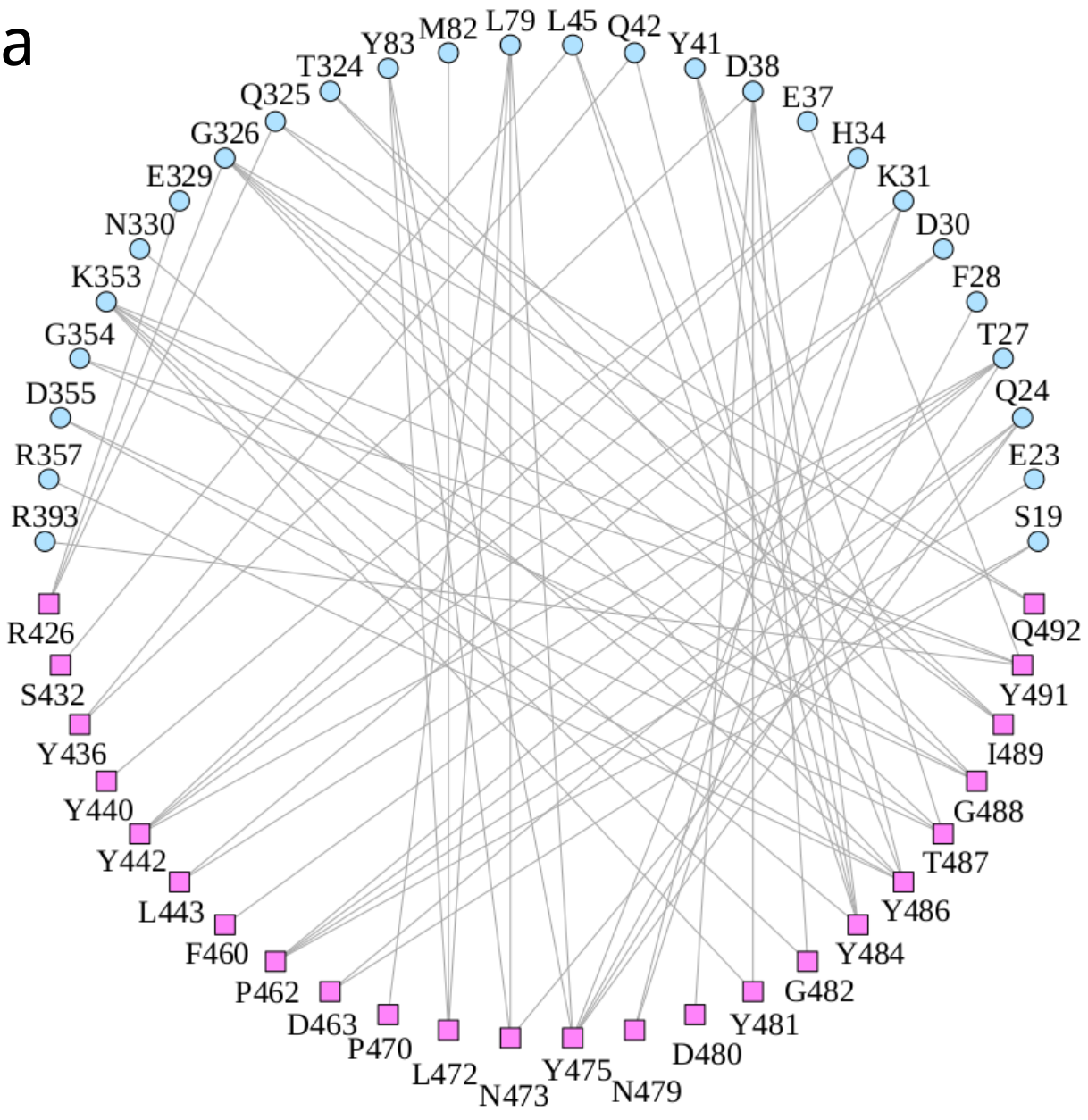}
\end{minipage}%
\hspace{0.6cm}
\begin{minipage}{.48\textwidth}
   \centering
   \includegraphics[angle=0,origin=r,width=9cm]{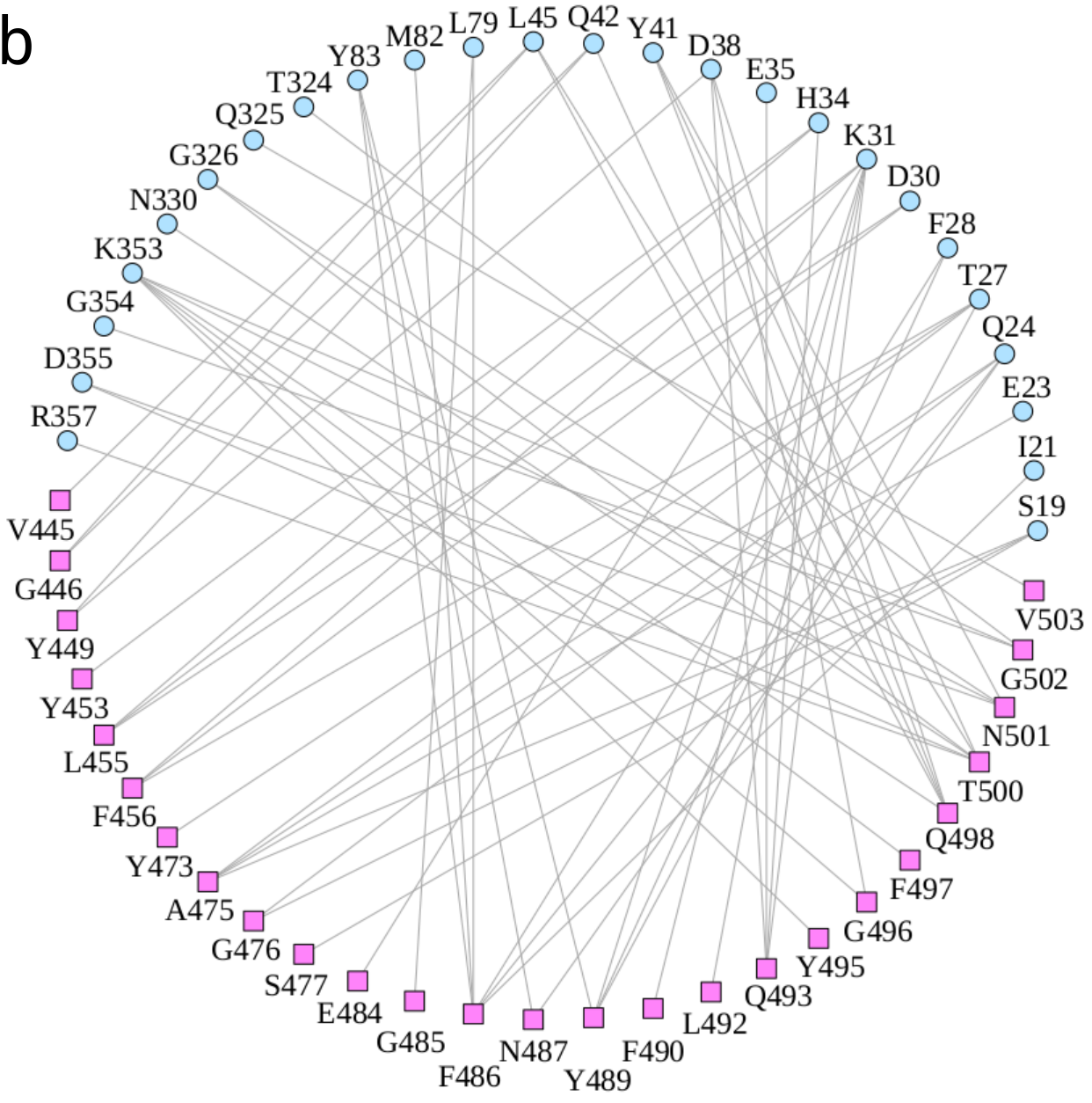}
\end{minipage}
\vspace{0.3cm}
\caption{Contact networks between (a) SARS-CoV residues, and 
    (b)  SARS-CoV-2 residues located in the RBM regions with ACE2.
   SARS-CoV and SARS-CoV-2 residues are in magenta color while human ACE2 residues 
   are in blue color.}
 \label{fig:network}
\end{figure*}

  Figure \ref{fig:network}a displays the residues of SARS-CoV RBM in direct contact
    with hACE2 as determined by the hydrophobic-hydrophilic properties
    of the interface residues as predicted by the CSU program \cite{CSU-1999}.
  This bipartite network of contacts highlights the importance of
    residues that are located near the YGF sequence and
contributes to the stabilization of SARS-CoV complexed with hACE2.
  For example, 
  Y475 makes hydrogen bond (H-B) contacts with Q24, F28, and Y83; 
  N479 with K31, and H34;
  Y486 with Y41, N330, and R357; and T487 with Y41.

\begin{figure*}
\begin{centering}
\includegraphics[width=18.0cm]{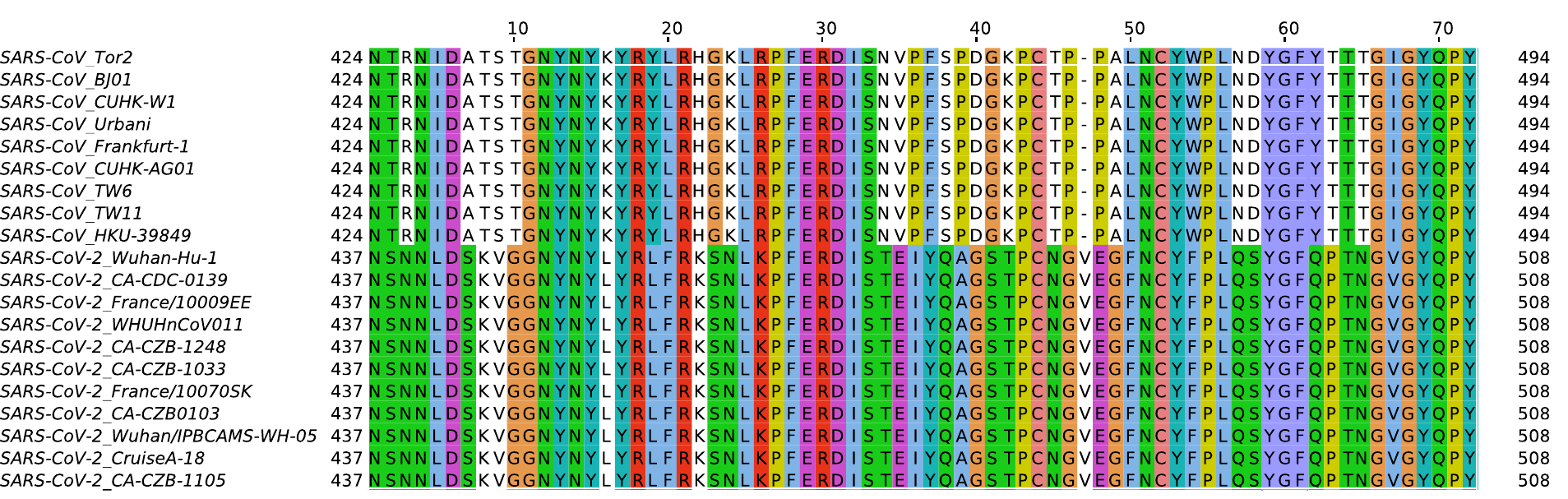}
\par\end{centering}
\caption{Sequence alignments of human CoVs restricted to RBM residues.
  The medium purple color highlights the YGFY pattern followed by 
  the mutation Y498Q in the RBM of SARS-CoV-2 strains.}
\label{fig:human}
\end{figure*}

    Figure~\ref{fig:pocket_sars}b 
displays the interface of 
\mbox{SARS-CoV-2} RBD spike-protein complexed with hACE2 (blue color).
    Now, the sequence YGFY observed in SARS-CoV is replaced by YGFQ as a result
    of sequence alignments shown in 
Fig~\ref{fig:human}.
    The single-point mutation Y484$\rightarrow$Q498 
 replaces a hydrophobic residue in SARS-CoV by a hydrophilic one in 
 SARS-CoV-2.
  
   Figure~\ref{fig:human}  
   compares residue sequences of human SARS-CoV and 
\mbox{SARS-CoV-2} strains aligned with RBM of SARS-CoV Tor2, an epidemic 
strain isolated from humans during the SARS epidemic in 2002-2003.
   The human Tor2 strain has high affinity for hACE2 \cite{cui-2019-nature}.
   We highlight in this figure 
in medium purple color the hydrophobic sequence
YGFY typical of SARS-CoV, occurring at positions 481-484 in the spike protein.
   The corresponding mutated sequence occurs now at positions 495-498 in
\mbox{SARS-CoV-2} spike protein. 

   The important residues for the interface interaction found in SARS-CoV are 
mutated in \mbox{SARS-CoV-2.}
  The sequence alignments show the mapping,
  Y442$\rightarrow$L455, 
  L472$\rightarrow$F486, 
  N479$\rightarrow$Q493,  
  D480$\rightarrow$S494, and  
  T487$\rightarrow$N501.     
  These mutations do not present a drastic
change in their hydrophobic character \cite{wimley-1996},
thus preserving the overall receptor-binding topological structure for 
these viruses.
    In particular, residues L455 and Q493 in SARS-CoV-2 preserve the noted 
favourable interactions with the residues E35 and K31 in hACE2 
\cite{yi-2020-Nature} 
   (Fig~\ref{fig:residues_sars}b). 
     Interestingly, a new GF sequence appears in the RBM of 
SARS-CoV-2 strains as a consequence of the mutation L472$\rightarrow$F486,
producing a small hydrophobic surface, but does not seem to disrupt
the proposed topological formation mechanism for ACE2 binding.
 No other GF sequence appears in their RBD.
  
    Figure \ref{fig:network}b displays the SARS-CoV-2 RBM residues in direct contact
    with hACE2 as predicted by the CSU program, showing again the importance of residues
    close to the hydrophobic pocket.
    Details of protein-protein binding interfaces can be
quite different among strains, likely related to their infectivity degree.
   It has been noted that mutations in RBM residue T487 in SARS-CoV have an important role 
in the human-to-human and animal-to-human transmission of SARS-CoV
\cite{cui-2019-nature,li-2015-review,qu-2005}.


\begin{table}[!htb]
\begin{tabular}{l cl} 
\toprule
                       & $\Delta \Delta G$  & ~~~ Effect on the  \\ 
   SARS-CoV            &  (kcal/mol)        & ~~~ complexation \\
\cmidrule(rr){1-1} \cmidrule(rr){2-3}
  N479E                &  0.76  & neutral\,$^{(2)}$ \\
  N479K                &  0.66  & neutral\,$^{(2)}$ \\
  N479Q                &  0.40  & neutral\,$^{(2)}$ \\
  N479R                &  0.08  & neutral\,$^{(2)}$ \\
  N479S                &  0.63  & neutral\,$^{(2)}$ \\
  G482D, Y484F         &  2.40  &  highly destabilizing\,$^{(2)}$\\
  G482D, Y484H         &  2.68  &  highly destabilizing\,$^{(2)}$ \\
  G482D, Y484N         &  3.24  &  highly destabilizing\,$^{(2)}$ \\
  G482D, Y484Q         &  2.47  &  highly destabilizing\,$^{(2)}$\\
  G482D, Y484T         &  3.26  &  highly destabilizing\,$^{(2)}$\\
  Y484F                &  0.64  & neutral\,$^{(**)}$  \\ 
  Y484H                &  1.57  &  highly destabilizing\,$^{(1)}$ \\
  Y484N                &  2.19  &  highly destabilizing\,$^{(1)}$ \\
  Y484Q                &  1.44  & neutral\,$^{(*)}$ \\
  Y484T                &  1.92  &  highly destabilizing\,$^{(1)}$ \\
  N479E, G482D, Y484F  &  4.58  &  highly destabilizing\,$^{(2)}$\\
  N479E, G482D, Y484H  &  4.90  &  highly destabilizing\,$^{(2)}$          \\
  N479E, G482D, Y484N  &  4.69  &  highly destabilizing\,$^{(2)}$           \\
  N479E, G482D, Y484Q  &  4.58  &  highly destabilizing\,$^{(2)}$          \\
  N479E, G482D, Y484T  &  4.67  &  highly destabilizing\,$^{(2)}$\\
  \\
 SARS-CoV-2            &        &  \\  
\cmidrule(rr){1-1}    \cmidrule(rr){2-3} 
  Q498Y                &  0.16  & neutral\,$^{(2)}$ \\
  G496D, Q498F         &  2.75  &  highly destabilizing\,$^{(2)}$\\
  G496D, Q498H         &  2.52  &  highly destabilizing\,$^{(2)}$\\
  G496D, Q498N         &  1.72  &  highly destabilizing\,$^{(1)}$ \\
  G496D, Q498T         &  1.68  &  highly destabilizing\,$^{(1)}$  \\
  Q493S                &  0.89  & neutral\,$^{(*)}$ \\
  Q493S, G496D, Q498F  &  3.63  &  highly destabilizing\,$^{(2)}$ \\
  Q493S, G496D, Q498H  &  4.33  &  highly destabilizing\,$^{(2)}$ \\
  Q493S, G496D, Q498N  &  2.96  &  highly destabilizing\,$^{(2)}$\\
  Q493S, G496D, Q498T  &  2.85  &  highly destabilizing\,$^{(2)}$\\
\bottomrule
\end{tabular}
\caption{Changes in binding affinity of human SARS-CoV and 
human SARS-CoV-2 spike RBD complexed with hACE2 upon mutation as 
predicted by MutaBind2 method.
Prediction effects are classified as low-confidence prediction: $(1)$, or  
high-confidence prediction: \mbox{$(2)$.}
Here we adopt the classification presented in MutaBind method \cite{mutabind} because 
of the high similarity between the ROC (receiver operating 
characteristic) curves in both methods.}
\label{tab:free}
\end{table}


   Now, we investigate the relevance of the hydrophobic pocket driven
by the YGFY sequence in promoting the stability of SARS-CoV spike receptor 
binding domain complexed with hACE2.
   To this end, we conducted a series of mutations to estimate the change in
binding affinity $\Delta \Delta G$ using the MutaBind2 method 
\cite{mutabind2}.
 
\begin{figure*}[!ht]
\begin{centering}
\includegraphics[width=18.0cm]{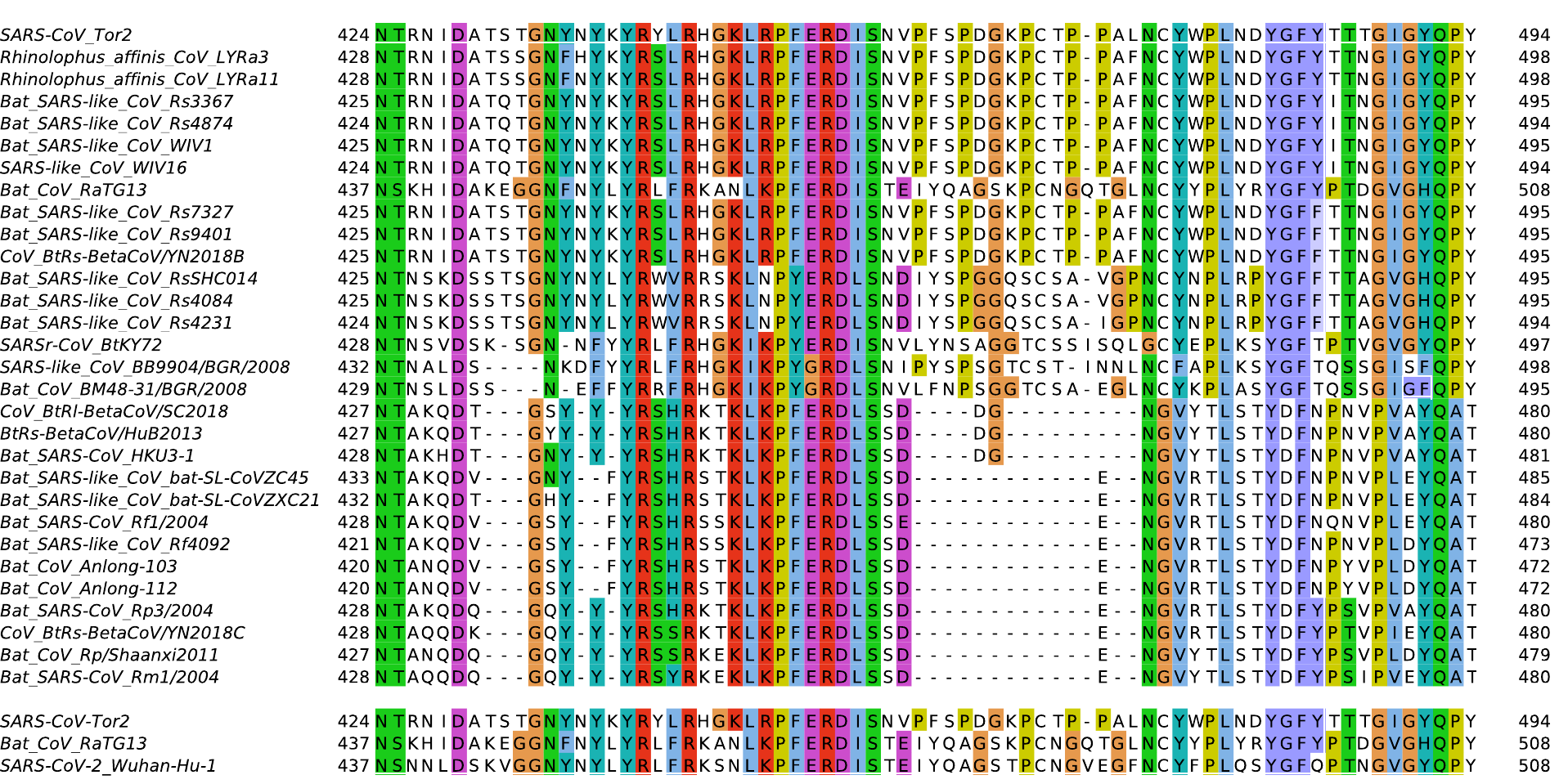}
\par\end{centering}
\caption{Sequence alignment of bat CoVs restricted to RBM residues
of SARS-CoV Tor2.
  The residues of YGFY pattern are in medium purple color.
  Last three alignments are placed together for direct amino acid sequence
   comparison.}
\label{fig:bat-clean}
\end{figure*}

\begin{figure*}
\begin{centering}
\includegraphics[width=18.0cm]{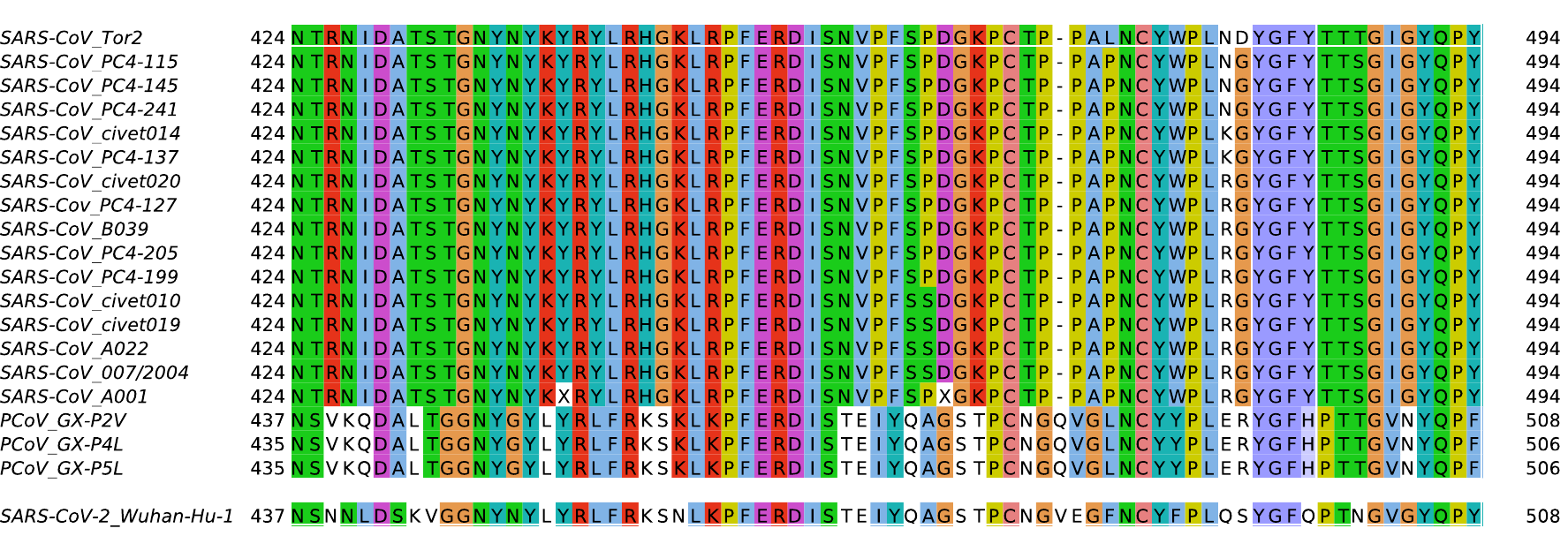}
\par\end{centering}
\caption{Sequence alignment of palm civet CoVs and pangolin PCoVs
restricted to RBM residues of SARS-CoV Tor2.
  The residues of YGFY pattern are in medium purple color.
  Last line includes the SARS-CoV-2 sequence for comparison.}
\label{fig:civet-pangolin}
\end{figure*}

   Initially, we investigate the influence of N479 mutation by the residues 
E, K, Q, R, and S on the complexation.
   The mutations 
   N479E, N479K, N479Q, N479R, and N479S have been observed, respectively
  in pangolin strains (see Fig. \ref{fig:civet-pangolin}),
     bat and palm civet strains (see Fig. \ref{fig:bat-clean}  and \ref{fig:civet-pangolin}),
     human SARS-CoV-2 (see Fig. \ref{fig:human}),
     bat and palm civet strains (see Fig. \ref{fig:bat-clean} and \ref{fig:civet-pangolin}), and
     bat strains (see Fig. \ref{fig:bat-clean}).
   The calculation of $\Delta \Delta G$  for these mutations
does not indicate any appreciable effect on SARS-CoV spike RBD/hACE2 binding 
affinity due to its small variation, as displayed in Table \ref{tab:free}.
   Therefore, we may conclude that N479 does not enhance the binding affinity 
of spike RBD to hACE2, and could as well be replaced by any of the above 
residues, preserving the hydrophilic character  \cite{wimley-1996}
and conformational stability.

   Specific mutations were also performed to investigate the importance of the  
residues forming the hydrophobic surface patch for the complexation stability.
   Residues Y481 and F483 are conserved through all SARS-CoV strains we have analysed in 
this work, while G482 and Y484 are mutated in some strains (bats and pangolins).
  Therefore, we mutated G482 and F484 by plausible residues,
i.e., the ones that occur in other strains to evaluated the changes in the binding 
affinity.
   To perform this double mutation, we fixed G482D, a mutation observed in bat strains, 
followed by
   Y484F, Y484N, and Y484T, observed in bats;
   Y484H, observed in pangolins; and
   Y484Q, observed in human SARS-CoV-2.
  The free-energy changes for these double mutations
strongly indicate the desestabilization of human SARS-CoV complexed with hACE2 (Table \ref{tab:free}).
  It is interesting to note that G482D 
 together with  Y484Q decreases the binding affinity in human SARS-CoV
because they decrease the hydrophobicity of the initial YGFY pocket.
  Table \ref{tab:free} also displays the changes in binding affinity for 
single mutations of Y484, a highy connected residue with hACE2 (see Fig. \ref{fig:network}).
  
   Next, we mutated N479 followed by mutations at sites 482  and 484
to analyse the consequences on the binding affinity of this triple mutation
by disrupting the hydrophobic surface patch.
   Again, we fixed, for example, the mutations N479E and G482D.
   The impacts of this set of mutations can be seen in Table \ref{tab:free}.
   What was considered to be a neutral mutation, N479E shows high 
destabilizing effect in the new conformational environment.
   Similar destabilizing effects on the human SARS-CoV spike RBD complexed
with hACE2 are obtained for the important residue T487 when one mutates
the  residues forming the hydrophobic pocket (data not shown).
  
  Now, we repeat the above  procedure to investigate the role
of residues YGFQ for human SARS-CoV-2 RBD/hACE2 binding affinity.
  The mutation Q498Y does not alter the binding affinity for the complexation
because the predicted $\Delta \Delta G$ is about 0.16 kcal/mol. 
  We also analysed the impact of double mutations in the YGFQ sequence
on the stability of the complexation, see Table \ref{tab:free}.
  To this end, we fixed, for example G496D and replaced Q498 
by the  residues F, H, N, and T  that appear in strains of other species.
  The predicted changes in binding affinity by mutations indicate  
  destabilization of new complexations.

\subsection{Spike receptor-binding motifs in bats}

     It is known that not all SARS-CoV strains isolated from bat hosts 
have exploited ACE2 as a cellular attachment.
     Therefore, the set of amino acid sequences displayed in 
  Fig~\ref{fig:bat-clean}
may exemplify the successful relation 
between virus evolution and the binding mechanism.
     This set highlights in medium purple color the preserved amino acid
residues in the sequence YGFY, characteristics of human SARS-CoV.
     For comparison, we also display CoV strains with mutations 
in that SARS-CoV pattern to explore the relation between
the hypothesized mechanism and the cell receptor recognition.

     It has been demonstrated that 
   LYRa11 \cite{letko-2020},
   Rs3367 \cite{ge-Nature-2013},
   Rs4874 \cite{hu-PLOS-2017},
   WIV1, and
   WIV16   \cite{letko-2020,yang-JV-2016},
have the capacity to use ACE2 for cell entry as well 
RaTG13, in line with our hypothesis.
   Also, the near single-point mutation Y$\rightarrow$F in the next six strains
Rs7327, Rs9401, YN2018B, RsSHC014, Rs4084, and Rs4231,
   does not interfere, as expected, in the attachment mechanism.
   This conclusion is supported  by cell entry studies for
   Rs7327 \cite{letko-2020,hu-PLOS-2017},
   Rs9401,
   RsSHC014,
   Rs4084, and
   Rs4231 \cite{hu-PLOS-2017}, because they are in a group that is likely to 
use the ACE2 receptor.
   This mutation replaces a hydrophobic residue by another one
with higher hydrophobicity, reinforcing the conformational topology
for binding with the receptor. 
 This single-point mutation Y$\rightarrow$F in human SARS-CoV produced
 a neutral effect in the binding hACE2 (Table \ref{tab:free}).

    We identified in the next group constituted by BtKY72, BB9904/BGR/2008, and
BM48-31/BGR/2008, respectively the mutations Y487T, Y488T, and  
Y485T, decreasing the initial hydrophobicity of the expected pocket in these
strains.
    It seems unlikely that this mutation and amino acid residue deletions 
associated to Tor2 RBM sequence affect the YGF-based 
attachment mechanism for BtKY72 and BB9904/BGR/2008.
  Unfortunately, there is no available experimental data concerning
their receptors.
    It is important to remark that the residue F492 in \mbox{BM48-31/BGR/2008}
produces another hydrophobic sequence IGF at residues 490-492 
  (Fig~\ref{fig:bat-clean}).
    We speculate that this double occurrence may disrupt the aforementioned 
mechanism because of indications that BM48-31/BGR/2008 does not interact, 
at least with human ACE2 \cite{letko-2020}.
   No other GF sequence occurs in the RBD of these strains. 
   
    Next CoV strains in Fig~\ref{fig:bat-clean}
    do not contain such specific YGF sequences of residues
in the RBM neither in their RBD.
    We find the two-letter sequence GF in Rf1/2004, but it is located in RBD 
and with GF surrounded by hydrophilic residues.
   Although we have considered only part of the sequences that better align 
with RBM of Tor2, it has been demonstrated that the spikes of 
HuB2013,
 HKU3,
CoVZC45,
CoVZXC21,
Rf1,
Rf4092,
 and
Shaanxi2011 do not use hACE2, a result that is not just 
a consequence of deletions at the RBD \cite{letko-2020}.
     Further support has been presented against HKU3 in using hACE2
\cite{gralinski-2020}.
     It seems unlikely that Rm1/2004 infects hACE2
because its  unfavourable binding free energy \cite{jaramillo-2020}.
    Another result concludes that Rp3 is unable of infect  hACE2
or even bat ACE2 \cite{hoffmann-PLOS-2013}.
 
    We have placed together the alignments involving Tor2, RaTG13, and 
SARS-CoV-2 at the end of 
   Fig~\ref{fig:bat-clean} 
  for further comparison.
    The whole genome of RaTG13 shares 96\%  amino acid sequence identity 
with SARS-CoV-2, and it is considered the most closely related genome to 
this CoV \cite{zhang-CurrBio-2020}.
   Considering its spike protein, and RBM, 
RaTG13 shares respectively 97\% and 76\% amino acid identity with SARS-CoV-2.
   For comparison, 
RaTG13 shares 79\%, 77\%, and 53\% identity, respectively, 
for the whole genome, spike protein, and RBM with SARS-CoV Tor2.
   Therefore, SARS-CoV-2 is mostly similar to RaTG13 than SARS-CoV strains
in all regions.

\subsection{Spike receptor-binding motifs in palm civets and pangolins}

    To explore further the role of YGF-based attachment mechanism,
we exhibit comparative residue sequences for civet and pangolins,
again aligned with RBM of SARS-CoV Tor2
  (Fig~\ref{fig:civet-pangolin}).
   This figure shows that the pattern  YGFY characteristic of 
human SARS-CoV is maintained for the collected data, but with a
single-point mutation Y$\rightarrow$H for pangolin hosts PCoV.
    It is worth to observe that even the shorter two-letter GF sequence 
is not found in the RBD of these strains, which could 
promote another hydrophobic pocket.

     We have included SARS-CoV-2 on the last line of 
  Fig~\ref{fig:civet-pangolin}
for a direct comparison.
     PCoV GX-P2V shares 79\%, 77\%, and 50\% amino acid identity with Tor2, 
respectively for whole genome, spike protein, and RBM aligned with Tor2.
     In relation to SARS-CoV-2, PCoV GX-P2V shares 85\%, 92\%, and 75\% 
amino acid identity, respectively for whole genome, spike protein, and RBM.
     It is believed that human SARS-CoV passed from palm civets to humans in the 
2002-2003 epidemic because their genome sequences are highly similar 
  \cite{cui-2019-nature,qu-2005,li-2008-JVI}.
     The amino acid alignments show an almost identical RBM between 
human SARS-CoV, 
represented by Tor2 strain, and collected data from palm civet strains. 
     This identification also includes the YGF-based mechanism  able to use 
ACE2 proteins.
     Nevertheless, 
     these alignments display high similarity between
pangolins and SARS-CoV-2, which also support previous conclusions on pangolins
being the probable origin of SARS-CoV-2 
\cite{zhang-CurrBio-2020,lam-2020-pangolin}.
     However, based on our data related to host receptor binding and their
RBM and S-protein alignments, we can not discard bat RaTG13-like strain as 
also the possible origin of SARS-CoV-2.

\subsection{SARS-CoV and hCoV-NL63: only functionally related}

    Although there is no many available experimental data identifying the 
viral receptor-binding protein for CoVs, it is well established that human 
SARS-CoV and hCoV-NL63 both employ ACE2 as the cell receptor to 
infect host cells  \cite{hoffmann-2005-NL63,hoffmann-2006-NL63}.
    Interestingly, SARS-CoV and hCoV-NL63 domains do not present high 
sequence similarity.  
   For example, their  spike-S1 subunities share only 10\% in similarity.
   Other features separate SARS-CoV and hCoV-NL63 \cite{milewska-2014}.
   SARS-CoVs are classified as $\beta$-{\it coronavirus} with subgenus 
{\it sarbecovirus},
while hCoV-NL63 is in genus $\alpha$-{\it coronavirus} and subgenus {\it setracovirus}.
   Although hCoV-NL63 also enters the cell via endocytosis, its functional
receptor requires heparan sulfate proteoglycans for the initial attachment, 
representing an important extra factor for ACE2 to act as a functional receptor 
\cite{milewska-2018,milewska-2014}.
    Moreover, the spike-S1 glycoprotein of SARS-CoV binds more
efficiently ACE2 than the corresponding  spike-S1 of NL63 (NL63-S)  \cite{glowacka-2010}.
    This may be linked to the fact that SARS-CoV and NL63-S contact ACE2
differently, a conclusion based upon the experimental results that  
NL63-S does not bind to ACE2 through a single and large domain  
 \cite{hoffmann-2006-NL63,wu-2009}.
   Actually, different RBD regions have been identified  within NL63-S.
   One of these regions was positioned at residues 476-616
and comprising three discontinuous RBM regions, RBM1 (residues 497-501), RBM2
(residues 530-540), and RBM3 (residues 575-594) 
\cite{li-2007-NL63,lin-2008-NL63,lin-2011-NL63}.
   A slightly different RBD has been identified for this CoV \cite{wu-2009}.
   It would be located at residues 482-602, also with three discontinuous RBM regions,
which surround a shallow cavity at hCoV-NL63-ACE2 binding interface.
   Curiously, its spike protein alignment with Tor2 does not 
show the expected residue pattern in the corresponding RBM 
of Tor2 nor in the aforementioned RBD regions of NL63-S.
   This may help to explain the unusual pathway of binding to ACE2 
for this CoV.

\section{Methods}

   We performed single and  multiple residue mutations to 
estimate the importance of the conserved residues in human SARS-CoV and 
SARS-CoV-2 RBM forming the hydrophobic pocket in establishing 
specific interactions with hACE2.
   Mutations may affect the spike receptor-binding complexed
with hACE2 either leading to higher, lower or even neutral binding affinity.
   Thus, we apply the fast and accurate MutaBind2 method \cite{mutabind2} to 
estimate the binding free-energy change 
 $\Delta \Delta G =   \Delta G^{\textrm{mut}} - \Delta G^{\textrm{wt}}$
 upon mutation to predict its functional effects.
 This method compares free-energy changes between 
 mutated and wild-type three-dimensional conformations.
  The binding free-energy change upon single mutation was also evaluated
  with the predictor BeAtMuSiC \cite{dehouck-music}
  based on a set of statistical potentials extracted from experimental
  mutational data. This computational method predicted very similar
  effects on the complexation (data not shown) as described in Table \ref{tab:free}.
\\
  
 \noindent
 {\bf Bioinformatic tools}
 
 We have also used the bioinformatic tools BLAST and ClustalW for 
sequence alignment and analysis of CoV strains, and 
 Jalview to examine and edit various sequence alignments.
  Figures showing the conformational complexations
  were prepared using PyMol.  
  The list of GenBank accession codes for the spike proteins analysed in 
this work is available in supplementary Table S1.

  \section{Conclusion}

    We have analysed a number of CoV strains to support the hypothesis 
that SARS-CoV and SARS-CoV-2 strains share a common evolutionary mechanism 
for the initial attachment to ACE2. 
    Moreover, we speculate that the YGF-based mechanism can act as a 
protein signature to distinguish CoVs able to use ACE2 as a cell entry 
receptor whenever this residue sequence is located at the CoV RBM region.
    For example, bat-SL-CoV ZC45 and ZXC21 are closely related sequences 
to  SARS-CoV-2 with overall genome identity of $\sim$ 89\% and 
can be promptly put under suspicious in their ACE2 binding affinity
because the lack of such signature.
    Of course, as exemplified by hCoV-NL63, we can not  discard that another
mechanism can act helping such ACE2 binding.
    It must be accentuated that the occurrence of other XGF sequences, mainly
with X being a hydrophobic residue, in the RBM, or even in the RBD region, 
can disrupt the proposed topological mechanism for ACE2 binding. 
    This because it might introduce hydrophobic loops promoting a new 
ligand-substrate recognition.

\section{Supporting information}

\paragraph*{Table S1}

{\bf GenBank accession numbers for the coronavirus sequences 
analysed in this study.}

\subsection*{Acknowledgements}
   
 P.P.D.C. and N.A.A. gratefully acknowledge financial supports from the 
 Brazilian agencies  CAPES,  and FAPESP, process 
 \mbox{2015/16116-3,} respectively.
\\
   
{\bf Competing interests}
   The authors declare no competing financial interests.


 \begin{figure*}
 \begin{centering}
 \includegraphics[width=20.0cm]{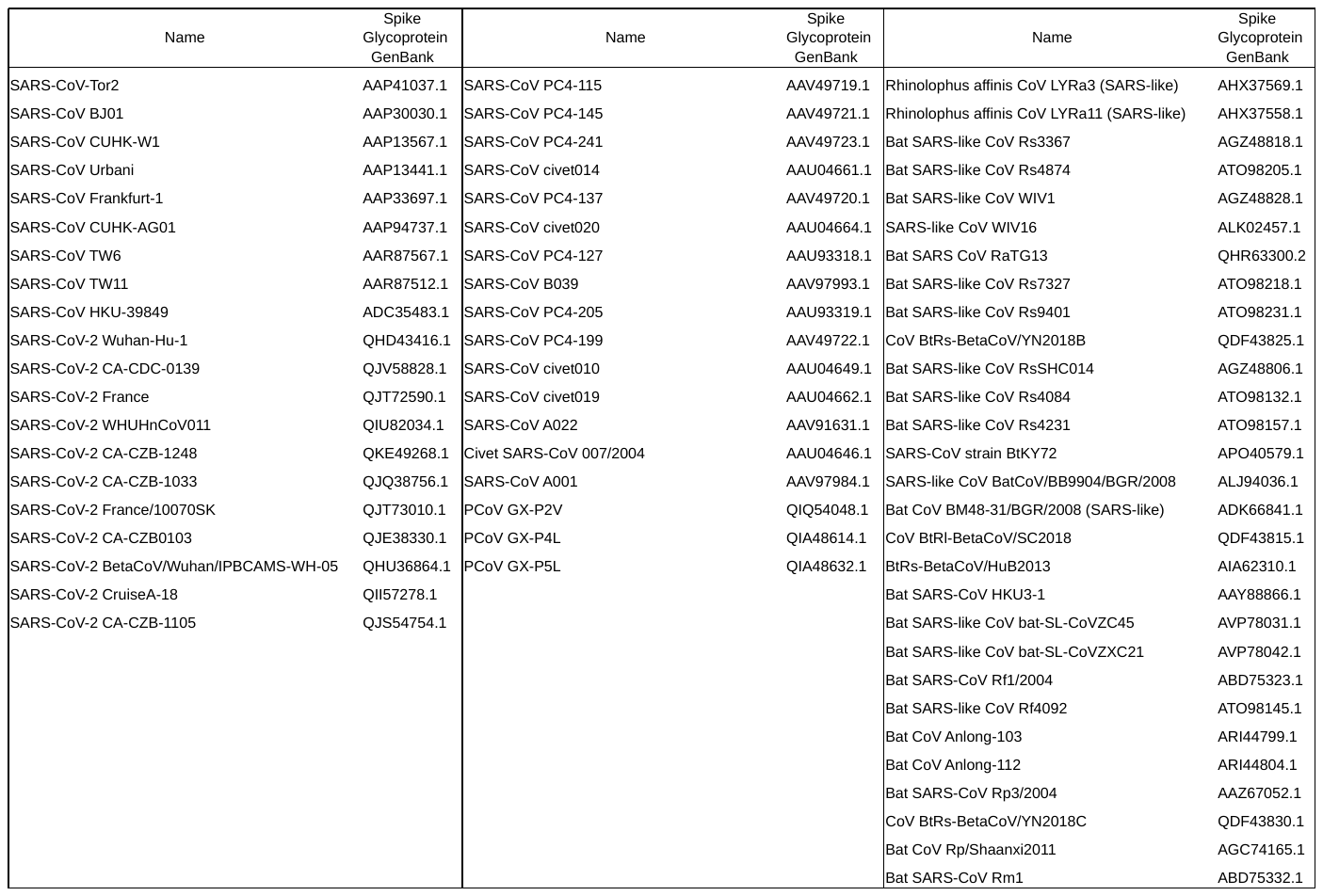}
 \par\end{centering}
 Table S1: GenBank accession numbers for the coronavirus sequences analysed in this study.
\label{fig:extra}
\end{figure*}


\begin{thebibliography}{99}

\bibitem{pan-2017} Pan, D., Nolan J., Williams K.H., Robbins M.J., 
    Weber K.A., 2017.
    Abundance and distribution of microbial cells and viruses in an alluvial aquifer. 
    Front Microbiol 8, 1199. 
    https://doi.org/10.3389/fmicb.2017.01199

\bibitem{shi-2018} Shi, M., Lin, X.-D., Chen, X., Tian, J.-H., Chen, L.-J., 
          Li, K., et al., 2018.
      The evolutionary history of vertebrate RNA viruses.
      Nature 556, 197-202. 
      https://doi.org/10.1038/s41586-018-0012-7
      
\bibitem{nasir-2014} Nasir, A., Forterre, P., Kim, K.M., 
      Caetano-Anoll\'es, G., 2014.
     The distribution and impact of viral lineages in domains of life.
     Front Microbiol  5, 194. 
     https://doi.org/10.3389/fmicb.2014.00194
     
\bibitem{koonin-2020} Koonin, E.V., Dolja, V.V., Krupovic, M., Varsani, A., 
     Wolf, Y.I., Yutin, N., et al., 2020.
     Global organization and proposed megataxonomy of the virus world. 
     Microbiol Mol Biol Rev  84, e00061-19. 
     https://doi.org/10.1128/MMBR.00061-19  

\bibitem{claverie-2006} Claverie, J.M., 2006.
    Viruses take center stage in cellular evolution.
    Genome Biol 7, 110. 
    https://doi.org/10.1186/gb-2006-7-6-110
      
\bibitem{sano-2004} Sano, E., Carlson, S., Wegley, L., Rohwer, F., 2004.
     Movement of viruses between biomes.
     Applied Environmental Microb 70, 5842-5846. 
     https://doi.org/10.1128/aem.70.10.5842-5846.2004
     
\bibitem{blerkom-2003}  Van Blerkom, L.M., 2003.
       Role of viruses in human evolution.
      Yearbook Phys Anthropology 46, 14-46. 
      https://doi.org/10.1002/ajpa.10384

\bibitem{filee-2003} Fil\'ee, J., Forterre, P., Laurent, J., 2003. 
      The role played by viruses in the evolution of their hosts: 
      a view based on informational protein phylogenies.
      Research Microbiology 154, 237-243. 
      https://doi.org/10.1016/S0923-2508(03)00066-4

\bibitem{koonin-2013} Koonin, E.V., Dolja, V.V., 2013.
      A virocentric perspective on the evolution of life.
      Curr Opinion Virol  3, 546-557. 
      http://doi.org/10.1016/j.coviro.2013.06.008

\bibitem{enard-2016} Enard, D., Cai, L., Gwennap, C., Petrov, D.A., 2016.
      Viruses are a dominant driver of protein adaptation in mammals.
      eLife  5, e12469. 
      http://doi.org/10.7554/eLife.12469.001
 
 \bibitem{barana-2003} Baranowski, E., Ruiz-Jarabo, C.M., Pariente, N., 
         Verdaguer, N., Domingo, E., 2003.
    Evolution of cell recognition by viruses: a source of biological novelty 
    with medical implications.
    Advances Virus Res 62, 19-111. 
    https://doi.org/10.1016/S0065-3527(03)62002-6
  
\bibitem{weng-2020}  Ng, W.M., Stelfox, A.J., Bowden, T.A., 2020.
     Unraveling virus relationships by structure-based phylogenetic classification.
     Virus Evol 6, veaa003.  
     https://doi.org/10.1093/ve/veaa003

\bibitem{ahola-2015} Ahola, T., Karlin, D.G., 2015.
     Sequence analysis reveals a conserved extension in the capping enzyme of the 
     alphavirus supergroup, and a homologous domain in nodaviruses.
     Biol Direct 10, 16. 
     https://doi.org/10.1186/s13062-015-0050-0
 
\bibitem{ICTV-2020} Gorbalenya, A.E., Baker, S.C., Baric, R.S., de Groot, R.J., 
     Drosten, C., Gulyaeva, A.A.,  et al., 2020.
     The species Severe acute respiratory syndrome-related coronavirus: 
     classifying 2019-nCoV and naming it SARS-CoV-2. 
     Nat Microbiol 5, 536-544. 
     https://doi.org/10.1038/s41564-020-0695-z
 
\bibitem{jaimes-2020} Jaimes, J.A., Andr\'e, N.M., Chappie, J.S., 
        Millet, J.K., Whittaker, G.R., 2020.
      Phylogenetic analysis and structural modeling of SARS-CoV-2 spike 
      protein reveals an evolutionary distinct and proteolytically 
      sensitive activation loop.
      J Mol Biol 432, 3309-3325. 
      https://doi.org/10.1016/j.jmb.2020.04.009

\bibitem{masters-2006}    Masters, P.S., 2006.
       The molecular biology of coronaviruses. 
       Adv Virus Res 66, 193-292. 
       https://doi.org/10.1016/S0065-3527(06)66005-3

\bibitem{cheng-2007} Cheng, V.C.C., Lau, S.K.P., Woo, P.C.Y., Yuen, K.Y., 2007.  
       Severe acute respiratory syndrome coronavirus as an agent of
       emerging and reemerging infection.
       Clinical Microbiol Rev 20, 660-694. 
       https://doi.org/10.1128/CMR.00023-07

\bibitem{su-2016} Su, S., Wong, G., Shi, W., Liu, J., Lai, A.C.K., 
      Zhou, J., et al., 2016. 
     Epidemiology, genetic recombination, and pathogenesis of coronaviruses. 
     Trends Microbiol 24, 490-502. 
     http://doi.org/10.1016/j.tim.2016.03.003

\bibitem{li-2016-review} Li, F., 2016.
       Structure, function, and evolution of coronavirus spike proteins.
       Annu Rev Virol 3, 237-261. 
       https://doi.org/10.1146/annurev-virology-110615-042301
   
\bibitem{sanjuan-2016} Sanju\'an, R., Domingo-Calap, P., 2016.
      Mechanisms of viral mutation.
      Cell Mol Life Sci 73, 4433-4448. 
      https://doi.org/10.1007/s00018-016-2299-6 
     
\bibitem{lau-2007} Lau, S.K.P., Woo, P.C.Y., Li, K.S.M., 
         Huang, Y., Wang, M., Lam, C.S.F., et al., 2007.
    Complete genome sequence of bat coronavirus HKU2 from chinese 
    horseshoe bats revealed a much smaller spike gene with a different 
    evolutionary lineage from the rest of the genome.
    Virol 367, 428-439. 
    https://doi.org/10.1016/j.virol.2007.06.009
  
\bibitem{woo-2012} Woo, P.C.Y., Lau, S.K.P., Lam, C.S.F., 
           Lau, C.C.Y., Tsang, A.K.L., Lau, J.H.N., et al., 2012.
    Discovery of seven novel mammalian and avian coronaviruses in the genus deltacoronavirus supports bat coronaviruses as the gene source of 
    alphacoronavirus and betacoronavirus and avian coronaviruses as the 
    gene source of gammacoronavirus and deltacoronavirus. 
    J Virol  86, 3995-4008. 
    http://doi.org/10.1128/JVI.06540-11
    
\bibitem{beniac-2006} Beniac, D.R., Andonov, A., Grudeski, E., 
       Booth, T.F., 2006.
    Architecture of the SARS coronavirus prefusion spike. 
    Nature Struct Mol Biol 13, 751-752. 
    http://doi.org/10.1038/nsmb1123

\bibitem{li-2012} Li, F., 2012. 
    Evidence for a common evolutionary origin of coronavirus
    spike protein receptor-binding subunits. 
    J Virol 86, 2856-2858. 
    http://doi.org/10.1128/JVI.06882-11

\bibitem{song-2018} Song, W., Gui, M., Wang, X., Xiang, Y., 2018. 
    Cryo-EM structure of the SARS coronavirus spike glycoprotein in 
    complex with its host cell receptor ACE2. 
    PLoS Pathog 14, e1007236. 
    https://doi.org/10.1371/journal.ppat.1007236
  
\bibitem{belouzard-2012} Belouzard, S., Millet, J.K., Licitra, B.N., 
           Whittaker, G.R., 2012.
    Mechanisms of coronavirus cell entry mediated by the viral spike protein. 
    Viruses 4, 1011-1033. 
    https://doi.org/10.3390/v4061011
    
\bibitem{hoffmann-2020} Hoffmann, M., Kleine-Weber, H., Schroeder, S., 
       Kr{\"u}ger, N., Herrler, T., Erichsen, S.,  et al., 2020.
       SARS-CoV-2 cell entry depends on ACE2 and TMPRSS2 and is blocked 
       by a clinically proven protease inhibitor.
       Cell 181, 271-280.
       https://doi.org/10.1016/j.cell.2020.02.052

\bibitem{letko-2020} Letko, M., Marzi, A., Munster, V., 2020.
      Functional assessment of cell entry and receptor
      usage for SARS-CoV-2 and other lineage B betacoronaviruses.
      Nature Microbiol 5, 562-569. 
      https://doi.org/10.1038/s41564-020-0688-y

\bibitem{walls-2020} Walls, A.C., Park, Y.J., Tortorici, M.A., Wall, A., 
               McGuire, A.T., Veesler, D., 2020.
    Structure, Function, and Antigenicity of the SARS-CoV-2 Spike Glycoprotein. 
    Cell 180, 281-292. 
    https://doi.org/10.1016/j.cell.2020.02.058.
  
\bibitem{wang-2008-cell} Wang, H., Yang, P., Liu, K., Guo, F., 
             Zhang, Y., Zhang, G., et al., 2008.
    SARS coronavirus entry into host cells through a novel
    clathrin- and caveolae-independent endocytic pathway.
    Cell Res 18, 290-301. 
    https://doi.org/10.1038/cr.2008.15

\bibitem{wang-2008-virus} Wang, S., Guo, F., Liu, K., Wang, H., 
           Rao, S., Yang, P., et al., 2008.
     Endocytosis of the receptor-binding domain of SARS-CoV spike protein
     together with virus receptor ACE2.
     Virus Research 136, 8-15. 
     https://doi.org/10.1016/j.virusres.2008.03.004
     
\bibitem{yuan-2017-nature} Yuan, Y., Cao, D., Zhang, Y., Ma, J., 
          Qi, J., Wang, Q., et al., 2017.
   Cryo-EM structures of MERS-CoV and SARS-CoV spike glycoproteins reveal 
   the dynamic receptor binding domains.
    Nature Commun 8, 15092. 
    https://doi.org/10.1038/ncomms15092

\bibitem{milewska-2018} Milewska, A., Nowak, P., Owczarek, K., Szczepanski, A., 
      Zarebski, M., Hoang, A., et al., 2018.
    Entry of human coronavirus NL63 into the cell. 
    J Virol 92, e01933-17. 
    https://doi.org/10.1128/JVI.01933-17
  
\bibitem{ou-2020} Ou, X., Liu, Y., Lei, X., Li, P., Mi, D., 
        Ren, L., et al., 2020.
     Characterization of spike glycoprotein of SARS-CoV-2 on virus entry 
     and its immune cross-reactivity with SARS-CoV. 
     Nature Commun 11, 1620.  
     https://doi.org/10.1038/s41467-020-15562-9
     
\bibitem{yang-2014-PNAS} Yang, Y., Du, L., Liu, C., Wang, L., Ma, C., 
           Tang, J., et al., 2014. 
     Receptor usage and cell entry of bat coronavirus HKU4
     provide insight into bat-to-human transmission of MERS coronavirus.
     Proc Natl Acad Sci USA 111, 12516-12521. 
     http://www.pnas.org/cgi/doi/10.1073/pnas.1405889111
 
\bibitem{li-2015-review} Li, F., 2015. 
     Receptor recognition mechanisms of coronaviruses: 
     a decade of structural studies. 
     J Virol 89, 1954-1964. 
     http://dx.doi.org/10.1128/JVI.02615-14
 
\bibitem{weiss-2005} Weiss, S.R., Navas-Martin, S., 2005.
      Coronavirus pathogenesis and the emerging pathogen severe acute
      respiratory syndrome coronavirus.
      Microbiol Mol Biol Rev 69, 635-664. 
      https://doi.org/10.1128/MMBR.69.4.635-664.2005
 
 \bibitem{cui-2019-nature}  Cui, J., Li, F., Shi, Z.-L., 2019. 
    Origin and evolution of pathogenic coronaviruses.
    Nature Rev Microbiol 17, 181-192. 
    https://doi.org/10.1038/s41579-018-0118-9
 
\bibitem{li-2005-embo} Li, W., Zhang, C., Sui, J., Kuhn, J.H.,
          Moore, M.J., Luo, S., et al., 2005. 
   Receptor and viral determinants of SARS-coronavirus adaptation 
   to human ACE2. 
   EMBO J 24, 1634-1643. 
   https://doi.org/10.1038/sj.emboj.7600640
 
\bibitem{wan-2020} Wan, Y., Shang, J., Graham, R., Baric, R.S., Li, F., 2020.
    Receptor recognition by the novel coronavirus from Wuhan: an analysis 
    based on decade-long structural studies of SARS coronavirus. 
    J Virol 94, e00127-20. 
    https://doi.org/10.1128/JVI.00127-20
    
\bibitem{li-2005-science309} Li, F., Farzan, M., Harrison, S.C., 2005. 
    Structure of SARS coronavirus spike receptor-binding domain 
    complexed with receptor. 
    Science 309, 1864-1868. 
    https://doi.org/10.1126/science.1113611

\bibitem{wu-2012} Wu, K., Peng, G., Wilken, M., Geraghty, R.J., 
          Li, F., 2012. 
     Mechanisms of host receptor adaptation by severe acute respiratory 
     syndrome coronavirus.
     J Biol Chem 287, 8904-8911. 
     https://doi.org/10.1074/jbc.M111.325803.

\bibitem{lan-2020-nature} Lan, J., Ge, J., Yu, J., Shan, S., Zhou, H., 
          Fan, S., et al., 2020.
     Structure of the SARS-CoV-2 spike receptor-binding domain bound 
     to the ACE2 receptor.
     Nature 581, 215-228. 
     https://doi.org/10.1038/s41586-020-2180-5.

\bibitem{he-2020} He, J., Tao, H., Yan, Y., Huang, S.-Y., Xiao, Y., 2020.
     Molecular mechanism of evolution and human infection with SARS-CoV-2.
     Viruses 12, 428. 
     https://doi:10.3390/v12040428
   
\bibitem{pickart-2004} Pickart, C.M., Cohen, R.E., 2004. 
     Proteasomes and their kin: proteases in the machine age. 
     Nature Rev Mol Cell Biol 5, 177-187. 
     https://doi.org/10.1038/nrm1336

\bibitem{finley-2009} Finley, D., 2009. 
    Recognition and processing of ubiquitin-protein conjugates by the proteasome. 
    Annu Rev Biochem 78, 477-513. 
    https://doi.org/10.1146/annurev.biochem.78.081507.101607
 
\bibitem{dieckmann-1998} Dieckmann, T., Withers-Ward, E.S., Jarosinski, M.A., 
        Liu, C.-F., Chen, I.S.Y., Feigon, J., 1998.
    Structure of a human DNA repair protein UBA domain that interacts
    with HIV-1 Vpr.
    Nature Struct Biol 5, 1042-1047.
    https://doi.org/10.1038/4220
    
\bibitem{wilkinson-2001} Wilkinson, C.R.M., Seeger, M., 
          Hartmann-Petersen, R., Stone, M., Wallace, M., 
           Semple, C., et al., 2001.
    Proteins containing the UBA domain are able to bind to multi-ubiquitin chains. 
    Nature Cell Biol 3, 939-943.
    https://doi.org/10.1038/ncb1001-939
   
\bibitem{bertolaet-2001} Bertolaet, B.L., Clarke, D.J., Wolff, M., 
       Watson, M.,  Henze, M., Divita, G., et al., 2001.
   UBA domains mediate protein-protein interactions between two DNA 
   damage-inducible proteins.
    J Mol Biol 313, 955-963. 
    https://doi.org/10.1006/jmbi.2001.5105

\bibitem{madura-2002} Madura, K., 2002.
    The ubiquitin-associated (UBA) domain: On the path from prudence to prurience
    Cell Cycle 1, 235-244. 
    https://doi.org/10.4161/cc.1.4.130
    
\bibitem{raasi-2005} Raasi, S., Varadan, R., Fushman, D., Pickart, C.M., 2005.
    Diverse polyubiquitin interaction properties of ubiquitin-associated domains. 
     Nature Struct Mol Biol 12, 708-714. 
     https://doi.org/10.1038/nsmb962

\bibitem{long-2008} Long, J., Gallagher, T.R.A., Cavey, J.R., 
     Sheppard, P.W., Ralston, S.H., Layfield, R., et al., 2008.
     Ubiquitin recognition by the ubiquitin-associated domain of p62 
     involves a novel conformational switch. 
     J Biol Chem 283, 5427-5440. 
     https://doi.org/10.1074/jbc.M704973200

\bibitem{tse-2011} Tse, M.K., Hui, S.K., Yang, Y., Yin, S.-T., 
     Hu, H.-Y., Zou, B., et al., 2011.
    Structural analysis of the UBA domain of X-linked inhibitor of apoptosis 
    protein reveals different surfaces for ubiquitin-binding and self-association. 
    PLoS ONE 6, e28511.
    https://doi.org/10.1371/journal.pone.0028511

 \bibitem{cabe-2018} Cabe, M., Rademacher, D.J., Karlsson, A.B., 
         Cherukuri, S., Bakowska, J.C., 2018.
    PB1 and UBA domains of p62 are essential for aggresome-like induced
    structure formation.
    Biochem Biophys Res Commun 503, 2306-2311.
    https://doi.org/10.1016/j.bbrc.2018.06.153
   
\bibitem{geetha-2002} Geetha, T., Wooten, M.W., 2002.
    Structure and functional properties of the ubiquitin binding protein p62.
    FEBS Lett 512, 19-24.
    https://doi.org/10.1016/S0014-5793(02)02286-X
 
 
\bibitem{mueller-2002} Mueller, T.D., Feigon, J., (2002).
   Solution structures of UBA domains reveal a conserved hydrophobic surface for
   protein-protein interactions.
     J Mol Biol 319, 1243-1255.

 
\bibitem{qu-2005}  Qu, X.-X., Hao, P., Song, X.-J., Jiang, S.-M., 
          Liu, Y.-X., Wang, P.-G., et al., 2005.
    Identification of two critical amino acid residues of the severe acute 
    respiratory syndrome coronavirus spike protein for its variation in 
    zoonotic tropism transition via a double substitution strategy.
     J Biol Chem 280, 29588-29595.
     https://doi:10.1074/jbc.M500662200

     
\bibitem{CSU-1999} Sobolev, V., Sorokine, A., Prilusky, J., Abola, E.E.,
   Edelman, M., 1999. 
   Automated analysis of interatomic contacts in proteins. Bioinformatics, 15, 327-332.

     
\bibitem{wimley-1996}  Wimley, W.C., White, S.H., 1996.
  Experimentally determined hydrophobicity scale for proteins 
  at membrane interfaces.
  Nature Struc Biol 3, 842-848.
  https://doi.org/10.1038/nsb1096-842

\bibitem{yi-2020-Nature} Yi, C., Sun, X., Ye, J., Ding, L., 
          Liu, M., Yang, Z., et al., 2020. 
   Key residues of the receptor binding motif in the spike protein of 
   SARS-CoV-2 that interact with ACE2 and neutralizing antibodies. 
   Cell Mol Immunol 17, 621-630.
   https://doi.org/10.1038/s41423-020-0458-z
   
\bibitem{mutabind2}
  Zhang, N., Chen, Y., Lu, H., Zhao, F., Alvarez, R.V., Goncearenco, A., 
  Panchenko, A.R., Li, M., 2020.
   MutaBind2: Predicting the Impacts of Single and Multiple Mutations on 
   Protein-Protein Interactions. iScience 23, 100939. 
 
\bibitem{mutabind}
  Minghui, L., Simonetti, F.L.,  Goncearenco, A.,  Panchenko, A.R., 2016.
  MutaBind estimates and interprets the effects of
  sequence variants on protein–protein interactions.
 Nucl Ac Res  44, W494-W501.

\bibitem{ge-Nature-2013} Ge, X.-Y., Li, J.-L., Yang, X.-L., 
      Chmura, A.A., Zhu, G., Epstein, J.H., et al., 2013.
     Isolation and characterization of a bat SARS-like coronavirus that 
     uses the ACE2 receptor. 
      Nature 503, 535-538.
     https://doi:10.1038/nature12711

\bibitem{hu-PLOS-2017} Hu, B., Zeng, L.-P., Yang, X.-L., 
        Ge, X.-Y., Zhang, W., Li, B., et al., 2017. 
    Discovery of a rich gene pool of bat SARS-related coronaviruses provides 
    new insights into the origin of SARS coronavirus. 
    PLoS Pathog  13, e1006698. 
    https://doi.org/10.1371/journal.ppat.1006698

\bibitem{yang-JV-2016} Yang, X.-L., Hu, B., Wang, B., Wang, M.-N., 
         Zhang, Q., Zhang, W., et al., 2016.
      Isolation and characterization of a novel bat coronavirus 
      closely related to the direct progenitor of severe 
      acute respiratory syndrome coronavirus. 
      J Virol 90, 3253-3256.
      https://doi:10.1128/JVI.02582-15

\bibitem{gralinski-2020}  Gralinski, L.E., Menachery, V.D., 2020.
   Return of the  Coronavirus: 2019-nCoV.
   Viruses 12, 135. 
   https://doi:10.3390/v12020135

\bibitem{jaramillo-2020}  Armijos-Jaramillo, V., Yeager, J., Muslin, C., 
      Perez-Castillo, Y., 2020.
   SARS-CoV-2, an evolutionary perspective of interaction with
   human ACE2 reveals undiscovered amino acids necessary for
   complex stability.
   bioRxiv preprint https://doi.org/10.1101/2020.03.21.001933

\bibitem{hoffmann-PLOS-2013}  Hoffmann, M., M\"uller M.A., Drexler, J.F., 
       Glende, J., 
       Erdt, M., G\"utzkow, T., et al., 2013. 
    Differential sensitivity of bat cells to infection by enveloped 
    RNA viruses: coronaviruses, paramyxoviruses, filoviruses, and 
    influenza viruses. 
    PLoS ONE 8, e72942.
    https://doi:10.1371/journal.pone.0072942

\bibitem{zhang-CurrBio-2020}  Zhang, T.,  Wu, Q., Zhang, Z., 2020. 
    Probable pangolin origin of SARS-CoV-2 associated with the COVID-19 outbreak. 
    Curr Biol 30, 1346-1351.
    https://doi.org/10.1016/j.cub.2020.03.022

\bibitem{li-2008-JVI} Li, F., 2008.
     Structural analysis of major species barriers between humans and
     palm civets for severe acute respiratory syndrome coronavirus infections.
     J Virol 82, 6984-6991.
     https://doi:10.1128/JVI.00442-08

 \bibitem{lam-2020-pangolin} Lam, T.T.-Y., Shum, M.H.-H., Zhu, H.-C., 
       Tong, Y.-G., Ni, X.-B., Liao, Y.-S., et al., 2020. 
   Identifying SARS-CoV-2-related coronaviruses in Malayan pangolins. 
   Nature 583, 282-285.
   https://doi:10.1038/s41586-020-2169-0

\bibitem{hoffmann-2005-NL63}  Hofmann, H., Pyrc, K., van der Hoek, L., 
          Geier, M., Berkhout, B., P\"ohlmann, S., 2005. 
    Human coronavirus NL63 employs the severe acute respiratory syndrome 
    coronavirus receptor for cellular entry.  
    Proc Natl Acad Sci USA 102, 7988-7993.
    https://doi:10.1073/pnas.0409465102

\bibitem{hoffmann-2006-NL63} Hofmann, H., Simmons, G., Rennekamp, A.J., 
      Chaipan, C., Gramberg, T., Heck, E., et al., 2006. 
   Highly conserved regions within the spike proteins of human 
   coronaviruses 229E and NL63 determine recognition
   of their respective cellular receptors.
   J Virol 80, 8639-8652.
   https://doi:10.1128/JVI.00560-06

\bibitem{milewska-2014} Milewska, A., Zarebski, M., Nowak, P., 
      Stozek, K., Potempa, J., Pyrc, K., 2014.
    Human coronavirus NL63 utilizes heparan sulfate proteoglycans for
    attachment to target cells. 
    J Virol  88, 13221-13230. 
    https://doi.org/10.1128/JVI.02078-14

\bibitem{glowacka-2010} Glowacka, I., Bertram, S., Herzog, P., Pfefferle, S., 
    Steffen, I., Muench, M.O., et al., 2010.
    Differential downregulation of ACE2 by the spike proteins of severe
    acute respiratory syndrome coronavirus and human coronavirus NL63.
    J Virol 84, 1198-1205.   
    https://doi:10.1128/JVI.01248-09

\bibitem{wu-2009}  Wu, K., Li, W., Peng, G., Li, F., 2009. 
    Crystal structure of NL63 respiratory coronavirus receptor-binding 
    domain complexed with its human receptor. 
    Proc Natl Acad Sci USA 106, 19970-19974.
    https://doi.org/10.1073/pnas.0908837106 
 
\bibitem{li-2007-NL63} Li, W., Sui, J., Huang, I.-C., Kuhn, J.H., 
       Radoshitzky, S.R., Marasco, W.A., et al., 2007.  
   The S proteins of human coronavirus NL63 and severe acute 
   respiratory syndrome coronavirus bind overlapping regions of ACE2. 
   Virology 367, 367-374.
   https://doi:10.1016/j.virol.2007.04.035

\bibitem{lin-2008-NL63} Lin, H.-X., Feng, Y., Wong, G., Wang, L., 
        Li, B., Zhao, X., et al., 2008. 
    Identification of residues in the receptor-binding domain (RBD) 
    of the spike protein of human coronavirus NL63 that are critical 
    for the RBD-ACE2 receptor interaction. 
    J Gen Virol   89, 1015-1024.
    https://doi:10.1099/vir.0.83331-0

\bibitem{lin-2011-NL63} Lin, H.-X., Feng, Y., Tu, X., 
        Zhao, X., Hsieh, C.-H., Griffin, L., et al., 2011.
   Characterization of the spike protein of human coronavirus NL63 in 
   receptor binding and pseudotype virus entry. 
   Virus Res 160, 283-293.
   https://doi:10.1016/j.virusres.2011.06.029
  
  
\bibitem{dehouck-music} Yves Dehouck, Y., Kwasigroch, J.M., Rooman, M., 
   Gilis, D., 2013. 
   BeAtMuSiC: prediction of changes in protein–protein binding affinity on mutations
   Nucleic Acids Res 41, W333-W339.
  
  
\end{thebibliography}
\end{document}